\documentclass[RNAAS]{aastex63}

\usepackage{amsmath}

\newcommand{\dd}{\textrm{d}}
\newcommand{\dif}{\textrm{d}}  % use "\dif x" for dx

\newcommand{\lfunc}{\phi}  % or \Phi or \varphi or...

\newcommand{\lpdf}{f}
\newcommand{\lcdf}{F}

\begin{document}

\title{The Break-By-One Gamma Distribution:\\
A Proper and Tractable Alternative to the Schechter Function\\
for Modeling Cosmic Populations}

\author[0000-0003-4692-4607]{Thomas J. Loredo}
\affiliation{Cornell Center for Astrophysics and Planetary Science\\
Cornell University\\
Ithaca, NY, USA 14853-6801}

\correspondingauthor{Thomas J. Loredo}
\email{loredo@astro.cornell.edu}

\begin{abstract}
The break-by-one gamma distribution has a probability density function resembling the Schechter function, but with the small-argument behavior modified so it is normalizable in commonly arising cases where the Schechter function is not.
Its connection to the gamma distribution makes it straightforward to sample from.
These properties make it useful for cosmic demographics.
\end{abstract}

% Useful for arXiv, but AAS will prompt for UAT concepts on submission.
% Star counts, lum funcs are in Astro techniques: An. math: Astro models.

\section{}

Many populations of astrophysical objects have key properties---e.g., luminosity, mass, or size---following power law distributions to a good approximation over a large range.
Often this is a consequence of self-similarity or scale invariance in the underlying physics, with the power law index reflecting aspects of that physics, and thus a key parameter of interest.
Self-similarity holds only over a finite range (e.g., constrained by nonlinearity); as a result, the power law behavior eventually ends, and the scale where that occurs is also an interesting parameter.
\citet[\S~4.4]{FB12-MSMA} provide an overview of the use of power law and modified power law distributions in astronomy.

A prototypical example is the galaxy luminosity function, $\lfunc(L)$, the number density of galaxies per unit volume and luminosity (I ignore distance or redshift dependence here for simplicity).
\citet{PS74-FormnGalSelfSimGravCond} described a self-similar gravitational condensation model predicting a distribution of galaxy masses that is a power law with a cutoff at high masses.
Motivated by this, \citet{S76-SchechterFunc} introduced a simple parametric model for galaxy (and cluster) luminosity functions, the widely-used \emph{Schechter function}.
The Schechter function is a power law that is smoothly truncated at large luminosities by an exponential decay factor:
\begin{equation}
\lfunc_S(L;L_*,\beta, A) =
  \frac{A}{L_*} \left(\frac{L}{L_*}\right)^{\beta} e^{-L/L_*},
\label{schechter-lf}
\end{equation}
where the parameters comprise a luminosity scale, $L_*$, a nominal low-luminosity power law index, $\beta$, and an amplitude parameter, $A$.
% There are varying conventions for parameterizing the amplitude of the Schecter function.
In this parameterization, $A$ has units of number density.
Formally, equation~\eqref{schechter-lf} implies a galaxy number density
\begin{equation}
n_S(L_*,\beta, A) = A \int \dif L 
  \frac{1}{L_*} \left(\frac{L}{L_*}\right)^{\beta} e^{-L/L_*},
\label{schecter-n}
\end{equation}
and a probability density function (PDF) for galaxy luminosities
\begin{equation}
\lpdf_S(L;L_*,\beta) = \frac{\lfunc_S(L;L_*,\beta, A)}{n_S(L_*,\beta, A)}
\label{schecter-pdf}
\end{equation}
(the amplitude parameter cancels on the RHS).
In some equivalent parameterizations, $A$ is often denoted $\lfunc_*$, although it neither has the units of $\lfunc$, nor is it equal to $\lfunc(L_*)$, as the symbol might suggest.
% Astronomers very typically fit such simple distributions to LF data.

% *** Mention use in salient context

The form of the Schecter function resembles that of the gamma distribution, a distribution for a nonnegative quantity, $x$, with a PDF given by
\begin{equation}
f_\Gamma(x;\alpha,s) =
  \frac{1}{s\Gamma(\alpha)} \left(\frac{x}{s}\right)^{\alpha-1} e^{-x/s},
\label{gamma-pdf}
\end{equation}
where $\alpha$ is the shape parameter, $s$ is the scale parameter, and $\Gamma(\cdot)$ denotes the gamma function.
To keep the gamma distribution proper (normalizeable), the parameters must satisfy $\alpha >0$, $s>0$.
The Schecter function would seem to imply a luminosity distribution that is proportional to a gamma distribution for $L$ (with shape parameter $\alpha = \beta - 1$ and scale parameter $s=L_*$).
However, when fit to equation~\eqref{schechter-lf}, the observed samples of many populations require $\beta \in (-2,-1)$, corresponding to $\alpha \in (-1,0)$, outside the range of validity for gamma distributions.
In this regime, equation~\eqref{schecter-n} gives an infinite number density, and the luminosity PDF in equation~\eqref{schecter-pdf} is thus undefined.

Very low luminosity sources are unobservable (due to detection thresholds reflecting the impacts of noise and background in observations), so in practice the \emph{observable} luminosity function is truncated at low luminosities, and the impropriety of the Schechter function is often ignored.
But the actual luminosity function must rise with decreasing luminosity less quickly than $L^{-\beta}$ (corresponding to $\beta$ becoming larger than $-1$), or be cut off at low luminosities (corresponding to there being a minimum galaxy luminosity).
For some populations, an increase in the power law index
% (i.e., flattening of the logarithmic slope)
is in fact observed at small values of the observable property.
For example, this is the case for the quasar luminosity function (e.g., \citealt{M+13-QuasarLumFunc}).
Similarly, the stellar initial mass function (related to the stellar luminosity function) has a low-mass (low-luminosity) index that flattens by $\approx 1$ (\citealt{K07-IMF-BPL}).
% (IMF models for most stellar samples typically do not include a Schechter-like exponential high-mass cutoff, but a cutoff is theoretically expected for large enough populations and is seen in some samples; see, e.g., \citealt{M00-HiMassIMF,T+11-UP2020-IMF}).

Motivated by these observations, 
% and by the utility of having a distribution resembling the Schechter function for both modeling and simulation studies, but proper when $\beta>-2$,  
I describe here a generalization of the Schecter function with $\lfunc \propto L^{\beta+1}$ at low luminosities, and thus integrable for $\beta > -2$.
Because of its close connection to the gamma distribution, I call it the \emph{break-by-one gamma distribution} (BB1Gamma).

A BB1Gamma luminosity PDF has three parameters: a mid-luminosity power law index, $\beta$, and two parameters defining the mid-luminosity range, $(l, u)$, with $l < u$ and  $u$ playing the role of $L_*$ in the Schecter function.
The power law index smoothly breaks to $\beta+1$ as $L$ decreases below $l$.
The BB1 luminosity PDF has the following functional form:
\begin{equation}
\lpdf_{\rm BB1}(L ; \beta,u,l) = 
  \frac{C(\beta,u,l)}{u}\left(1-e^{-L/l}\right) \left(\frac{L}{u}\right)^{\beta} e^{-L/u},
\label{lumPDF} 
\end{equation}
where the normalization constant $C(\beta,u,l)$ is
\begin{equation}
C(\beta,u,l) =
  \begin{cases} \dfrac{1}{\Gamma(\beta+1)\cdot\left(1-\frac{1}{\left(1+\frac{u}{l}\right)^{\beta+1}}\right)} 
    & \quad \text{if } \beta > -2\text{ and }\beta \ne -1; \\
 \dfrac{1}{\log\left(1+\frac{u}{l}\right)} & \quad \text{if } \beta=-1.
  \end{cases}
\label{normLumPDF} 
\end{equation} 
% A BB1Gamma distribution is a kind of mixture of two gamma distributions (one with negative weight) with the same shape parameter but different scale parameters, parameterized in a slightly clever way.
Note that as $l\rightarrow 0$, the BB1Gamma distribution becomes a gamma distribution (if $\beta > -1$).
%Also, in our computational implementation, the condition $\beta=-1$ of the first case is $-1.001<\beta<-0.999$.
I devised the BB1 distribution to have smooth power law break behavior at low $L$, yet also have an analytical normalization constant; it is proper for $\beta > -2$.
Thanks to its close connection to the gamma distribution, one can generate samples from the BB1 distribution using a straightforward modification of a widely-used algorithm for sampling from the gamma distribution due to \citet{AD74-SampGammaBetaPoisBinom}.
These properties make it useful for simulation studies.

A BB1Gamma luminosity function may be defined simply by multiplying the BB1 luminosity distribution by the galaxy spatial number density, $n$:
\begin{equation}
\lfunc_{\rm BB1}(L ; n,\beta,u,l) = n \lpdf_{\rm BB1}(L ; \beta,u,l).
\label{lumPDF} 
\end{equation}
Such a straightforward parameterization is not possible for the Schechter function because of its impropriety for typically observed values of $\beta$.

\begin{figure}[t]
\centerline{\includegraphics[width=\textwidth]{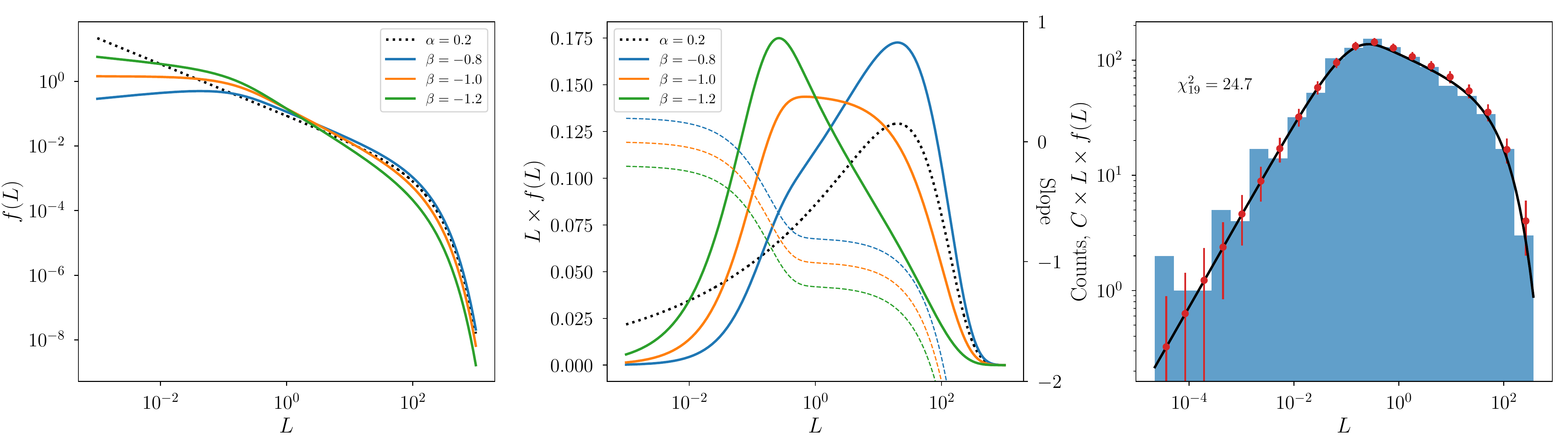}}
\caption{
Example BB1Gamma PDFs.
\emph{Left:} Three examples (solid lines) with $u=100$, $l=0.1$ (in generic units), and $\beta\in (-0.8, -1, -1.2)$, on log-log axes.
A gamma PDF is shown as a dashed line.
\emph{Middle:} The same cases, with the ordinate showing the PDF for $\log(L)$.
Dashed curves (right axis) show logarithmic slope.
\emph{Right:} Histogram shows random samples from the $\beta=-1.2$ case, in 20 bins of equal log width; ordinate shows counts, and $L f(L)$ scaled to give expected counts in such bins.
Red dots and error bars show predicted counts and standard deviations.
Pearson's $\chi^2$ for this sample is $24.7$ with 19 degrees of freedom.}
\label{bb1-3panes}
\end{figure}

Figure~\ref{bb1-3panes} shows example BB1Gamma PDFs, all with $u=100$ and $l=0.1$ (in generic units), with $\beta \in (-0.8, -1, -1.2)$ ($-0.8$ is a proper value for the Schechter function, corresponding to a gamma PDF with $\alpha=0.2$).
A gamma PDF is also shown for comparison.
The left panel uses conventional log-log axes.
The middle panel has a logarithmic abscissa, but uses $L f(L)$ (the PDF for $\log(L)$)as the ordinate; geometric area on this plot is proportional to (integrated) probability, so it more fairly depicts where samples from the distribution come from.
The dashed curves (right axis) show the local power law index, corresponding to the slope, $G(L)$, in $\log$-$\log$ space,
\begin{equation}
	G(L) \equiv \frac{\dd\log{\lpdf}}{\dd\log{L}} = \frac{L}{\lpdf} \frac{\dd \lpdf}{\dd L} = g(L) + \beta - \frac{L}{u},
\end{equation}
with
\begin{equation}
	g(L) = \frac{L}{l}\cdot\frac{1}{e^{L/l} - 1}.
\end{equation}
Evidently, $g(L) \rightarrow 0$ for $L \gg l$ and $g(L) \rightarrow 1$ for $L \ll l$.
Thus the logarithmic slope, $G(L)$, corresponds to an exponential cutoff at large $L$, and at small $L$, a slope of $\beta + 1$.
When $u\gg l$, there is a range where $L\gg l$ but $L\ll u$, and the logarithmic slope is $\approx \beta$ in that range.
The right panel shows a histogram of samples drawn from the $\beta=-1.2$ case.%
\footnote{See \url{https://github.com/tloredo/bb1gamma} for Python code producing this figure.}

Finally, the BB1 cumulative distribution function is 
\begin{equation}
\lcdf(L ; \beta,u,l) = 
  C(\beta,u,l)
  \left[ \Gamma(\beta + 1) - \gamma\left(\beta + 1, L/u\right) - \frac{\Gamma(\beta + 1)-\gamma\left(\beta + 1, L\cdot\left(\frac{1}{u}+\frac{1}{l}\right)\right)}{\left(1 + \frac{u}{l}\right)^{\beta + 1}} \right],
\label{lumCDF} 
\end{equation}
where $\gamma(\cdot, \cdot)$ denotes the upper incomplete gamma function.
\citet{S+18-CUDAHM} use the BB1Gamma distribution in simulation studies demonstrating luminosity function inference accounting for measurement error and selection effects for population sizes $\sim 10^6$, using a graphics processing unit (GPU).

\acknowledgments
This material is based upon work supported by the National Science Foundation under Grant No.~AST-1814840.

% Force a break here, else refs overlap the footnote.
\pagebreak

% \bibliography{BB1GammaNote}

\end{document}